\documentclass[utf8]{FrontiersinHarvard} 

\usepackage{url,hyperref,microtype,subcaption,graphicx}
\usepackage[onehalfspacing]{setspace}
\usepackage[]{color}

\newcommand{\beq}{\begin{equation}}
\newcommand{\seq}{\end{equation}}

\newcommand{\gv}[1]{\ensuremath{\mbox{\boldmath$ #1 $}}} 
\newcommand{\pd}[2]{\frac{\partial #1}{\partial #2}}
\newcommand{\pdtext}[2]{\partial #1/\partial #2}
\renewcommand{\div}[1]{\gv{\nabla} \cdot #1} 
\newcommand{\f}{\frac}  

\def\cv{\ensuremath{c_{\rm v}}}
\def\Js{\ensuremath{J_{\rm s}}}
\def\Jh{\ensuremath{J_{\rm h}}}
\def\Ers{\ensuremath{E_{r, {\rm s}}}}
\def\Erh{\ensuremath{E_{r, {\rm h}}}}

\defcitealias{Waters19a}{WP19}

\usepackage{letltxmacro,amsmath}
\LetLtxMacro{\originaleqref}{\eqref}
\renewcommand{\eqref}{Eq.~\originaleqref}

\def\keyFont{\fontsize{8}{11}\helveticabold }
\def\firstAuthorLast{Waters \& Proga} 
\def\Authors{Tim Waters\,$^{1,*}$ and Daniel Proga\,$^{2, 3}$}
\def\Address{$^{1}$Theoretical Division, Los Alamos National Laboratory, NM 87545, USA\\
$^{2}$Department of Physics \& Astronomy, University of Nevada, Las Vegas, Las Vegas, NV, 89154-4002, USA\\
$^{3}$Nevada Center for Astrophysics (NCfA), University of Nevada, Las Vegas, NV, USA}
\def\corrAuthor{Tim Waters}

\def\corrEmail{waters@lanl.gov}

\begin{document}
\onecolumn
\firstpage{1}

\title[Saturation mechanism of TI]{The saturation mechanism of \\thermal instability}

\author[\firstAuthorLast ]{\Authors} 
\address{} 
\correspondance{} 
\extraAuth{}

\maketitle

\begin{abstract}
The literature on thermal instability (TI) reveals that even for a simple homogeneous plasma, the nonlinear outcome ranges from a gentle reconfiguration of the initial state to an explosive one, depending on whether the condensations that form evolve in an isobaric or nonisobaric manner.  
After summarizing recent developments on the linear and nonlinear theory of TI, here we derive several general identities from the evolution equation for entropy that reveal the mechanism by which TI saturates:
whenever the boundary of the instability region (the Balbus contour) is crossed, a dynamical change is triggered that causes the comoving time derivative of the pressure to change sign.  This event implies that the gas pressure force reverses direction, slowing the continued growth of the condensation.  For isobaric evolution, this `pressure reversal' occurs nearly simultaneously for every fluid element in the condensation and a steady state is quickly reached.  For nonisobaric evolution, the condensation is no longer in mechanical equilibrium and the contracting gas rebounds with greater force during the expansion phase that accompanies gas reaching the equilibrium curve.  The cloud then pulsates because the return to mechanical equilibrium becomes wave-mediated.  We show that both the contraction rebound event and the subsequent pulsation behavior follow analytically from an analysis of the new identities.  Our analysis also leads to the identification of an isochoric TI zone and makes it clear that unless this zone intersects the equilibrium curve, isochoric modes can only become unstable if the plasma is in a state of thermal nonequilibrium.

\tiny
 \keyFont{ \section{Keywords:} thermal instability, plasma instabilities, nonadiabatic flows, multiphase gas dynamics, radiation hydrodynamics} 
\end{abstract}

\section{Introduction} 

Thermal instability (TI) is a linear instability of the equations of nonadiabatic gas dynamics that was first identified by \cite{Parker53}. He discussed an application to the solar atmosphere, namely the brightness fluctuations in solar flares and prominences, and there has recently been a resurgence of interest by the solar physics community in understanding the role of TI in prominence formation and the coronal rain phenomenon 
\citep[e.g.][]{Soler12,Antolin20,ClaesKeppens21,Soler22,Antolin22}.  TI became a topic of importance to the wider astrophysics community with the classic work by \cite{Field65}.  In recent years, various review articles on galactic outflows, the circumgalactic medium (CGM), and active galactic nuclei (AGN) have made it clear that `multiphase gas dynamics' has become recognizable as an astrophysical subfield of its own \citep[e.g.][]{Tumlinson17,Zhang18Review,Veilleux20,Laha20,FGO23,Choudhur23}.  

Use of the term `multiphase' warrants clarification in any given context, however.  In the original two-phase \citep{Field69} and three-phase \citep{McKee77} model of the interstellar medium (ISM), 
each phase is true to the sense of the word, representing a cold atomic or molecular phase interacting with a warm partially ionized phase.  This is the meaning of the term also in observations of molecular outflows from AGN, as well as in galactic winds and other studies related to the ISM/CGM.  Outside central cluster galaxies in the intracluster medium, gas temperatures are nearly virialized and `multiphase gas' refers to a multi-temperature plasma, one in which both `phases' are near collisional ionization equilibrium (CIE) conditions, the lower temperature plasma undergoing stronger bremstrahlung cooling and coronal line emission compared with the fully ionized plasma.  This is likewise the sense of the term on subparsec scale regions of AGN, albeit the plasma is much closer to photoionization equilibrium than to CIE due to the presence of an intense ionizing radiation field \citep[e.g.][]{Krolik81,Lepp85}.

In almost all cases where these astrophysical environments are modeled using hydrodynamical simulations that include radiative heating/cooling and multiphase gas appears, its production is attributable to TI reaching the nonlinear regime.  The dynamics associated with this regime first involves the saturation of TI, whereby the gas pressure force halts the exponential growth of condensation modes traversing a TI zone --- the location in density/temperature parameter space satisfying the generalized stability criteria first derived by 
\cite{Balbus86}.  As we review in \S{2} below, in addition to the entropy mode, there can be two isochoric condensation modes that are associated with a different TI zone than the entropy mode.  There is also the possibility that acoustic modes can become overstable within this isochoric TI zone, but it remains to be demonstrated that this can actually lead to multiphase gas production in global simulations.  
In this paper, we focus on understanding the process by which individual entropy modes saturate in a homogeneous plasma, although the equations used for this purpose apply to inhomogeneous flows and to the other modes as well.  These equations and our analysis of the saturation process is given in \S{3}, where we also discuss several topics that might benefit from a similar analysis.  In \S{4}, we summarize our results and address a couple controversial claims in the literature. 

\section{Summary of (non)linear theory results}
In the early literature on TI, emphasis is placed on isobaric and isochoric instability criteria rather than on the linear modes obeying these criteria.  Upon considering the full parameter space of TI, it becomes important to draw a distinction between the stability criteria, the individual modes, and the timescales dictating the type of nonlinear evolution.  This understanding will aid the presentation of our new results.  Here we attempt to clarify these concepts by summarizing our results from \citeauthor{Waters19a} (\citeyear{Waters19a}; hereafter \citetalias{Waters19a}).
In that paper, we revisited the original analysis of the governing cubic dispersion relation presented by \citet{Field65} to identify the nonisobaric regime of TI, and we also showed that at sufficiently long wavelengths there can be both a fast and slow isochoric condensation mode. 

In \citetalias{Waters19a}, we neglected to point out an inconsistency between two approaches to deriving the stability criterion of these isochoric modes under circumstances where the background flow is out of thermal equilibrium (i.e., for $\mathcal{L} \neq 0$, where $\mathcal{L}$ is the net sum of radiative heating and cooling processes).  Reworking Field's analysis after taking $\mathcal{L} \neq 0$, we found that his original criterion for isochoric instability is recovered, while as shown by \cite{Balbus86}, a different criterion follows from a direct perturbation analysis of the entropy equation.  By contrast, the dispersion relation analysis recovers Balbus' criterion in the case of the entropy mode.  As discussed in \S{2.4}, this discrepancy has implications for plasmas in states of thermal nonequilibrium (TNE), as Balbus' criterion implies the existence of an isochoric TI zone that would otherwise have gone undetected by linear theory.   

\subsection{Condensation modes}  
Upon linearizing the equations of nonadiabatic gas dynamics with perturbations of the form $\exp(\omega\,t + i\gv{k}\cdot\gv{x})$, where $\omega = \omega_R + i\omega_I$ is the complex-valued frequency of a mode with wavenumber $k=|\gv{k}|$, two speeds emerge: the phase velocity $v_p = \omega_I/k$ that gives the propagation speed of the mode, and the `condensation velocity' $v_c = \omega_R/k$ characterizing the speed of the advective flow that grows or damps the amplitude of the density perturbation.  Condensation modes are non-propagating ($v_p = 0$) linear modes with $v_c \neq 0$, whereas adiabatic acoustic modes are non-condensating ($v_c = 0$) with $v_p \neq 0$.
This statement holds so long as all nonadiabatic source terms aside from a heat flux due to thermal conduction can be written as a volumetric term $\mathcal{L} = \mathcal{L}(\rho,T)$.  

As discussed in detail in \S{2.3}, the classifier `isochoric' refers to the type of derivative of $\mathcal{L}/T$ that governs the stability of the isochoric modes; it does not, however, imply evolution that differs from the entropy mode during the linear growth phase. 
In a homogeneous gas (or when the local approximation holds), all three modes grow or damp according to the analytic solution 
\beq
\begin{split}
\rho(\gv{x},t) &= \rho_0 + A \rho_0 e^{\omega_R t} \cos(\gv{k}\cdot{\gv{x}})  \\
v(\gv{x},t) &= v_0 - A v_c e^{\omega_R t}\sin(\gv{k}\cdot{\gv{x}})  \\
p(\gv{x},t) &= p_0-A \rho_0 v_c^2 e^{\omega_R t} \cos(\gv{k}\cdot{\gv{x}}),
\end{split}
\label{eq:linear_profiles}
\seq
where $\rho_0$, $v_0$, and $p_0$ are the density, velocity, and pressure of the uniform background flow, $A$ is the perturbation amplitude, and $v_c$ is the solution to the cubic dispersion relation, which can be expressed as
\beq 
v_c = -\f{N_\rho}{k\,t_{\rm cool}} \f{R_\lambda + (v_c/c_{s,0})^2}{1 + (v_c/c_{s,0})^2}.
\label{eq:vcDR}
\seq
Here, $c_{s,0} = \sqrt{\gamma p_0/\rho_0}$ and $t_{\rm cool}$ is the characteristic cooling time defined as 
\beq
t_{\rm cool} \equiv \f{\mathcal{E}_0}{\Lambda_0},
\label{eq:tcool}
\seq  
where $\mathcal{E}_0 = \cv T_0$ is the gas internal energy 
and $\Lambda_0$ is the cooling rate (in units of ${\rm erg~s^{-1}~g^{-1}}$) 
evaluated in the background flow. 
The quantity $R_\lambda$ is the ratio 
\beq
R_\lambda \equiv \f{N_p}{\gamma N_\rho},
\label{eq:Rdef}
\seq
where $N_p$ and $N_\rho$ are dimensionless measures of how the net cooling rate varies with temperature:\footnote{The subscripted `$T=T_0$' here corrects a notational error in Eq.~6 of WP19; this bracketed term is equal to $(\pdtext{\mathcal{L}}{T})_p - \mathcal{L}/T$ evaluated at $T=T_0$.}
\beq
N_p = \f{T_0}{\Lambda_0}\left[T \left(\pd{\mathcal{L}/T}{T}\right)_p\right]_{T=T_0} + \left(\f{\lambda_F}{\lambda}\right)^2;
\label{eq:Np}
\seq 
\beq
N_\rho = \left.\f{T_0}{\Lambda_0}\left(\pd{\mathcal{L}}{T}\right)_\rho\right\vert_{T=T_0} + \left(\f{\lambda_F}{\lambda}\right)^2.
\label{eq:Nrho}
\seq 
The subscript on $R_\lambda$ emphasizes the wavelength dependence of this ratio; 
in \eqref{eq:Np} and \eqref{eq:Nrho}, $\lambda_F$ is the Field length defined as 
\beq
\lambda_F = 2\pi \sqrt{\f{\kappa_0\, T_0}{\rho_0 \Lambda_0}},
\seq
where $\kappa_0$ is the (isotropic) thermal conductivity evaluated in the background flow.

Condensation modes can be classified by which solution branch they occupy when solving \eqref{eq:vcDR}.  
Equivalently, they can be classified by their stability criteria.  The stability of the entropy mode 
is always governed by the sign of $N_p$.  In particular, when thermal conduction is negligible 
(requiring either $\lambda_F \rightarrow 0$ or $\lambda \gg \lambda_F$ in \eqref{eq:Np} and \eqref{eq:Nrho}), 
the stability of entropy modes depends only on the sign of the isobaric temperature derivative 
of $\mathcal{L}/T$, i.e., they are unstable if the inequality 
\beq 
\left(\pd{\mathcal{L}/T}{T}\right)_p  < 0,
\label{eq:BalbusCriterion}
\seq 
known as Balbus' instability criterion \citep[after][]{Balbus86}, is satisfied. 
Likewise, the stability of the fast/slow isochoric modes is determined 
by the sign of $N_\rho$; the instability criterion for them is
\beq 
\left(\pd{\mathcal{L}}{T}\right)_\rho  < 0
\label{eq:FieldCriterion}
\seq 
when neglecting thermal conduction.  Note the temperature derivative is not of $\mathcal{L}/T$ in the isochoric case; this is the discrepancy mentioned above that we will examine further in \S{2.4}.

Typically, only one of the condensation modes, namely the entropy mode or the fast isochoric mode, 
will exist in the plasma for any given value of 
\beq
R \equiv R_\lambda(\lambda_F = 0), 
\nonumber
\seq
the other two modes being acoustic.  
The main exception is the long wavelength regime for gas with $R < 0$: there is a critical 
wavelength $\lambda_{\rm crit}$ such that for $\lambda > \lambda_{\rm crit}$, the acoustic modes transition 
into the fast and slow isochoric condensation modes, so all three modes can exist simultaneously.  
This critical wavelength is, to a good approximation,
\beq 
\lambda_{\rm crit} = \f{2\pi}{|N_\rho|}\sqrt{-\f{B}{R}} \lambda_{\rm cool},
\seq 
where $B = 27(R-1/3)^2/4 - 1$ and $\lambda_{\rm cool}$ is the so-called cooling length defined as 
\beq
\lambda_{\rm cool} = c_{s,0}\,t_{\rm cool}.
\seq
For $\lambda < \lambda_{\rm crit}$, only the entropy mode exists; when the isochoric modes appear 
at $\lambda > \lambda_{\rm crit}$, they are stable when the entropy mode is unstable 
(i.e., when $N_p < 0$) and vice versa.  The $R>0$ regime is very different, as the fast isochoric mode takes the place of the entropy mode as the main condensation mode accompanying acoustic modes, and these acoustic modes are also unstable 
(or to use proper terminology, they are overstable), their stability being governed by the isochoric criterion 
provided $R<1$.  Note by \eqref{eq:Rdef} that instability when $R>0$ requires $N_p<0$ 
and $N_\rho < 0$ --- both the isobaric and isochoric instability criteria are satisfied.  For $R>1$, 
the only condensation mode that can exist is the fast isochoric one; the overstability of acoustic modes 
is no longer determined by the sign of $N_\rho$ but rather by the sign of the isentropic derivative 
$(\pdtext{\mathcal{L}}{T})_s$, where $s$ is the specific entropy.

\subsection{Isobaric versus nonisobaric regimes}
A number of properties can be inferred from \eqref{eq:linear_profiles} and \eqref{eq:vcDR}, making it clear under what circumstances a transition away from isobaric growth rates implies nonisobaric evolution in the nonlinear regime, regardless of which stability criterion is satisfied.  Notice that the quantity $A\,v_c\,e^{\omega_R\,t} $ is the maximum magnitude of the \textit{instantaneous} velocity at which plasma is supplied to the slightly cooler gas by the slightly warmer gas to grow the density perturbation with time.  At the end of the linear regime when $A\,e^{\omega_R\,t} \sim 1$, $v_c$ is the maximum velocity reached.  It is therefore the characteristic velocity of advective flows upon entering the saturation phase of nonlinear growth.  Stated differently, not only is the quantity $v_c$ the solution to the dispersion relation of the linearized equations of gas dynamics, but its value reveals what type of dynamics to expect in the nonlinear regime of TI. 

Our reference point is therefore the isobaric value of $v_c$ that can be derived by taking the limit $|v_c| \ll c_{s,0}$ in \eqref{eq:vcDR}; namely, $v_c = -N_p/(\gamma t_{\rm cool} k)$. Since $v_c \equiv \omega_R/k$, this corresponds to the growth rate $\omega_R = -N_p/(\gamma t_{\rm cool})$ derived by \cite{Field65} for the short wavelength limit (when neglecting thermal conduction) that defines the isobaric regime.  Unstable growth requires $\omega_R > 0$, giving $N_p < 0$ as the instability criterion.  This derivation of the maximum growth rate and Balbus' instability criterion from taking the limit $|v_c| \ll c_{s,0}$ thus reveals the dynamics associated with isobaric evolution: at the end of the linear regime, only tiny pressure gradients are necessary to give rise to flows with $|v_c| \ll c_{s,0}$ that regulate the growth and saturation of entropy modes.

Thus, the nonisobaric regime can be defined as the wavelengths for which solutions to \eqref{eq:vcDR} do not satisfy $|v_c| \ll c_{s,0}$.  As mentioned above, the entropy mode is the only condensation mode that can exist for wavelengths $\lambda < \lambda_{\rm crit}$ and $R<0$.  It occupies the solution branch of \eqref{eq:vcDR} with the property that $v_c$ rises smoothly from 0 in the small wavelength ($k\rightarrow \infty$) limit to $|v_c| = c_{s,0}\sqrt{-R}$ in the long wavelength ($k\rightarrow 0$) limit.  Hence, as the wavelength increases, entropy modes undergo a transition from isobaric to nonisobaric dynamics unless $|R| \ll 1$.  The two isochoric condensation mode branches only appear for $\lambda > \lambda_{\rm crit}$ and asymptote to $|v_c| = c_{s,0}\sqrt{-R}$ and $|v_c| = (k\,t_{\rm cool})^{-1}N_\rho$ in the $k\rightarrow 0$ limit.  We named the mode with $v_c \propto k^{-1}$ the fast isochoric mode due to this divergent property.  Because $k$ can always be chosen small enough to make $|v_c| > c_{s,0}$, this mode is associated with an explosive saturation regime at sufficiently long wavelengths, the expected outcome being a self-fragmentation of the cloud after the saturation phase.  With $|v_c|/c_{s,0} = \sqrt{-R}$ for both the entropy mode and slow isochoric mode in the $k=0$ limit, the advective flows of long enough wavelength modes will become supersonic during saturation provided $R \lesssim -1$.  In WP19, we speculated that this could also lead to self-fragmentation, referring to either instance of saturation leading to self-fragmenation as `splattering'.  This has since been demonstrated by \cite{GronkeOh20}, although they did not specify if the unstable condensation modes in their simulations were fast-isochoric ones or entropy modes, and they incorrectly attributed the self-fragmentation seen to the now discredited \citep[see WP19;][]{Das21,Farber22} `shattering' hypothesis of \cite{McCourt18} that is discussed below.

\subsection{Nonisobaric versus isochoric evolution}
The nonisobaric behavior at wavelengths giving $|v_c| \gtrsim 0.1 c_{s,0}$ but before reaching the splattering regime at $|v_c| \sim c_{s,0}$, has been confirmed by other researchers, and is the following.  The response to the halting of the compression phase once cooling gas reaches its equilibrium temperature is for the core of the condensation to `bounce' and enter an expansion phase.  \cite{JenningsLi} coined the term `contraction rebound' to conceptualize this saturation dynamics.
Further compressive/expansive motions recur as damped oscillations, causing the cloud to `pulsate', the descriptor used by \cite{GronkeOh20}.  If this rich dynamics is studied using a single entropy mode for the initial conditions, the frequency of the pulsations is the linear theory growth rate, $\omega_R$ 
\citepalias[see][]{Waters19a}.

In the literature on nonlinear TI, there are a number of conflicting claims about the differences between the short and long wavelength regimes and their relationships to isobaric and isochoric instability \citep[e.g.,][]{Meerson96,BurkertLin,Vazquez03,McCourt18,Mandelker21}. 
The most controversial claim came from \cite{McCourt18}, who presented simulations appearing to show that an isochorically cooling background flow containing unstable perturbations can spontaneously shatter during the nonlinear saturation phase of TI.  Their study was aimed at establishing $\lambda_{\rm cool}$ \textit{evaluated in the cold phase gas after TI saturates} as the more relevant length scale characterizing cloud sizes compared with the already established value, namely $\lambda_{\rm cool}$ evaluated in the background flow out of which the clouds condensed \citep[e.g.,][]{Perry85,BurkertLin}.  We showed in \citetalias{Waters19a} that $\lambda_{\rm cool}$ in the cold phase gas will generically be smaller than the Field length of the background flow, making \cite{McCourt18}'s hypothetical `cloudlets' subject to immediate evaporation \citep[see][]{Begelman90}.  Additionally, we interpreted their simulations as being a demonstration not of `shattering' but rather of `isobaric takeover', a process described succinctly by \cite{BurkertLin}: ``the fluctuations
that can first reach nonlinearity would dominate the growth of all perturbations with longer wavelengths and homogenize disturbances with smaller wavelengths. Thus, they determine the characteristic size and mass of the cold dense clumps that would emerge from the cooling of an initially nearly homogeneous cloud.''  In other words, since \cite{McCourt18} introduced a spectrum of perturbations into their initial conditions (a pre-existing cloud with a density contrast $\chi = 10$), the short wavelength condensation modes with the highest growth rates saturated while the longest wavelength ones were still in the linear regime, making it appear as though the cloud as a whole underwent fragmentation.  We return to this point in \S{4.1}.

To quote again from \citet{BurkertLin}, when discussing the nonisobaric regime, they state:
``Because of its long sound-crossing timescale, the perturbation cannot be compressed significantly while cooling; it cools almost isochorically (Parker 1953).''
We view this wording as confusing on account of the fact that, as already mentioned, all condensation modes undergo compression according to \eqref{eq:linear_profiles} during the linear growth phase, regardless of the sound-crosssing timescale.
To clarify matters, we note that there are two common usages of the word \textit{isochoric} in the literature on TI: (i) reference to the isochoric instability criterion or to the two condensation modes obeying this criterion; and (ii) reference to long timescales for any small amplitude thermal fluctuations in the background flow (which can be decomposed into a superposition of sinusoidal condensation modes of various wavelengths) to lead to changes in the density, i.e., for $t_\rho \equiv |\pdtext{\ln\rho}{t}|^{-1} \gg t_{\rm cool}$. 
Because $t_\rho$ can only be as large as the relevant dynamical time --- the perturbation sound crossing time, $t_{\rm cross}$, in local simulations --- usage (ii) entails having $t_{\rm cool} \ll t_{\rm cross}$.  This in turn requires distinguishing between short and long wavelength condensation modes, but the concept of `long wavelength' only makes sense when the modes become nonlinear so that they can excite and interact with sound waves to `be informed' of their length.  Once this interaction occurs, then if $t_{\rm cool} \ll t_{\rm cross}$, sound waves cannot quickly communicate the pressure changes accompanying thermal fluctuations. 
As we demonstrated in \citetalias{Waters19a}, what happens instead of isochoric evolution (the density remaining approximately constant in this nonlinear interaction phase with sound waves) is the characteristic nonisobaric behavior described above: oscillations commence throughout the condensation.
Thus, `isochoric evolution' is not a meaningful concept unless the oscillations fully dampen, this mechanical equilibrium state being reached only if the condensation is free of thermal disturbances. 

In summary, in the linear phase of TI, the descriptors `isobaric' and `isochoric' only have meaning as classifiers of the instability criteria.  When discussing evolution in the nonlinear phase of TI, `isobaric', `nonisobaric', and `isochoric' are associated with $t_{\rm cross} \lesssim t_{\rm cool}$, $t_{\rm cross} \gtrsim t_{\rm cool}$, 
and $t_{\rm cross} \gg t_{\rm cool}$, respectively.  
However, it is clear that the regime $t_{\rm cross} \gg t_{\rm cool}$ represents `extreme nonisobaric behavior' that will not allow $t_\rho \sim t_{\rm cross}$.  
For this reason, we feel that in studies discussing the long-wavelength regime of TI, phrases such as `the evolution should be nearly isochoric' should be interpreted as `the evolution should be highly nonisobaric'.

\subsection{TI zones}
All of the possibilities for TI can be assessed graphically given the function $\mathcal{L}$.  In general, this function can only be computed numerically using, for example, a photoionization code, and so it is impossible to arrive at analytic expressions for the instability thresholds --- the zero contours of $N_p$ and $N_\rho$.  
Recently, however, in extending the theory of TI to radiation hydrodynamics (RHD), we arrived at a simple function $\mathcal{L}$ that does have analytic expressions for $N_p$, $N_\rho$, and the curves defining their zero contours \citep[see][]{Proga22}.  In Appendix~A, we take the optically thin limit of these equations to facilitate their use in hydro/MHD codes and to make our results here easily reproducible.  In figure~1, we plot the contour $\mathcal{L} = 0$ on a pressure-density phase plane (referred to as just the `phase diagram' hereafter).  In textbook presentations of TI, typically only the regions where $\mathcal{L} > 0$ and $\mathcal{L} < 0$ are marked, masking any connection to nonlinear dynamics.  We will see below how overplotting TI zones makes the phase diagram useful for understanding nonisobaric evolution.  In \S{3}, we employ a combined graphical and analytical analysis to uncover the saturation process.

\begin{figure}[h!]
\begin{center}
\includegraphics[width=0.98\textwidth]{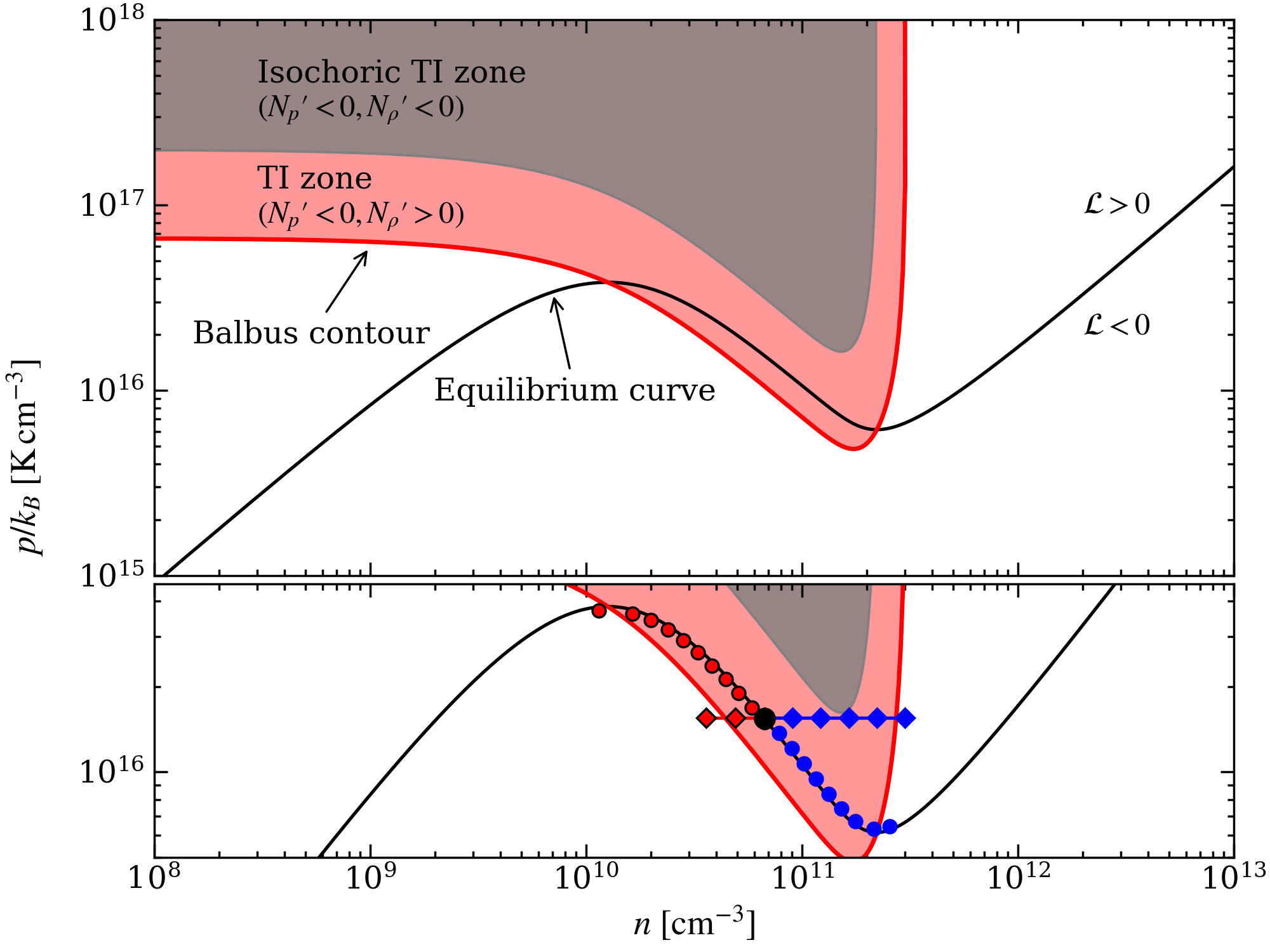}
\end{center}
\caption{Pressure-density phase diagram depicting the equilibrium curve (black line), the TI zone (pink region), the isochoric TI zone (gray region), and the boundary of the TI zone (the Balbus contour, red line).  In the bottom panel, we plot an initial unstable equilibrium state (black dot) along with two sets of `tracks' to depict the two limiting ways condensation modes begin to evolve in the nonlinear phase of TI, with blue (red) symbols denoting gas being cooled (heated): 
(i) nearly isobarically along horizontal tracks (diamonds);
(ii) highly nonisobarically along tracks following the equilibrium curve (dots).  The saturation mechanism begins at the location where the tracks end; it is triggered when the tracks cross the Balbus contour into a region of stability.}
\label{fig:1}
\end{figure}
 
We can use the cooling function defined in \eqref{eq:mathcalL} to highlight the significance of the discrepancy mentioned at the beginning of \S{2}, namely the existence of an isochoric TI zone according to Balbus' criterion and the lack of unstable parameter space for isochoric modes according to Field's criterion.  Taking the temperature derivative at fixed density of \eqref{eq:mathcalL} gives
\beq 
\left(\pd{\mathcal{L}}{T}\right)_\rho  =  \f{1}{2} c\,A_{\rm ff} \rho\, T^{-4.5}\left(aT^4 + 7 \Ers\right)
 + c A_C E_r. 
 \label{eq:L_Trho}
\seq 
Here, $E_r$ and $E_{r,s}$ are both constant values of the radiation energy density (see \eqref{eq:Erdef}) and $A_{\rm ff}$ and $A_C$ are positive constants.  We see that this derivative is always positive and thus cannot satisfy Field's instability criterion as given by \eqref{eq:FieldCriterion}.  However, Balbus' instability criterion for isochoric modes is $[\pdtext{(\mathcal{L}/T)}{T}]_{\rho} < 0$, which expands to
\beq 
\left(\pd{\mathcal{L}}{T}\right)_\rho < \f{\mathcal{L}}{T}.
\label{eq:BalbusCriterion_p}
\seq 
Substituting \eqref{eq:L_Trho} into \eqref{eq:BalbusCriterion_p} yields an expression that can be satisfied, demonstrating how allowing for $\mathcal{L} > 0$ increases the parameter space for TI.  We define any circumstance where the background flow has $\mathcal{L} \neq 0$ as it being in a state of thermal nonequilibrium (TNE).  There has been confusion in the literature regarding the onset of TI in a TNE state, which we address directly in \S{4.2}.  

At this point, it is useful to recognize that the contour where $(\pdtext{\mathcal{L}}{T})_{\rho} = \mathcal{L}/T$ is equivalent to the zero contour of the dimensionless quantity
\beq 
N_\rho' = \f{T_0}{\Lambda_0}\left[ T\left(\pd{\mathcal{L}/T}{T}\right) \right]_\rho.
\label{eq:Nrho_prime}
\seq 
The region on the phase diagram satisfying \eqref{eq:BalbusCriterion_p}, or equivalently $N_\rho' < 0$, is what we refer to as the isochoric TI zone; the gray region in figure~1 is therefore where the fast/slow isochoric modes are unstable and where acoustic modes can be overstable.   
The TI zone marking where entropy modes are unstable is likewise defined by the zero contour of the quantity 
\beq 
N_p' = \f{T_0}{\Lambda_0}\left[ T\left(\pd{\mathcal{L}/T}{T}\right) \right]_p.
\label{eq:Np_prime}
\seq 
We avoided terming this the `isobaric TI zone' because long wavelength entropy modes that give rise to nonisobaric evolution in the nonlinear regime of TI are still nevertheless governed by the `isobaric' instability criterion $N_p' < 0$.  Note that, upon neglecting thermal conduction, \eqref{eq:Np_prime} is identical to \eqref{eq:Np} (after evaluating it at a given $T_0$), whereas \eqref{eq:Nrho_prime} is equivalent to \eqref{eq:Nrho} only for $\mathcal{L}=0$.

From figure~1, notice that the gray region is a subset of the red region, implying both $N_p' < 0$ and $N_\rho' < 0$ there; in terms of the discussion from \S{2.1}, along the equilibrium curve this net cooling function lacks parameter space where $R < 0$ due to $N_p > 0$ and $N_\rho < 0$.  Again, the only way the isochoric TI zone will be entered is if gas is undergoing cooling with $\mathcal{L} > 0$, i.e. if the background flow is in a TNE state.  

The bottom panel of figure~1 illustrates our point about it being necessary to replace the concept of `isochoric evolution' with `highly nonisobaric evolution'.  The black dot marks an unstable position on the equilibrium curve. If the background flow is assigned initial conditions at this location, then the linear growth phase of TI takes place on this dot because any noticeable deviation away from the initial conditions implies nonlinear amplitudes.  The two sets of `tracks' shown therefore represent paths that can be taken by an unstable condensation mode once its amplitude becomes nonlinear.  The horizontal set corresponds to isobaric evolution.  A vertical set would correspond to isochoric evolution.  What is shown instead is what actually happens in the long-wavelength regime: a condensation mode follows the equilibrium curve.  This was first revealed by numerical simulations (see WP19), but it is clear in hindsight that this is what the nonlinear dynamics requires; when $t_{\rm cool} \ll t_{\rm cross}$, the gas will be driven to equilibrium as it evolves.\footnote{To view an animation comparing isobaric and nonisobaric evolution, please refer to the published article data \href{https://iopscience.iop.org/article/10.3847/1538-4357/ab10e1\#apjab10e1f5}{here}.}
 
 The red solid line in this figure is the boundary of the TI zone that we refer to as the Balbus contour.  We terminate the tracks once they cross this contour to emphasize that this stage of evolution marks the beginning of the saturation process, as we show in the next section.

\section{The saturation mechanism} 
The dynamics of how TI saturates is highly nonlinear, hence most prior work is based on constructing simplified dynamical models \citep[see][]{Meerson96} or analyzing the results of numerical simulations.
In \citetalias{Waters19a}, we associated the temporal event of the cooling gas `landing on the equilibrium curve' with the dynamical response henceforth referred to as `contraction rebound', to adopt the term introduced by \citet{JenningsLi}.  The sequence of events leading up to contraction rebound is initiated well before the cloud reaches the equilibrium curve.  As we show in \S{3.2}, the trigger is a change in sign of $N_p'$, i.e., when the gas crosses the boundary of the TI zone and thus switches from being unstable to stable. 

\subsection{Equations governing the nonlinear regime of TI}
Our starting point is the evolution equation for the specific entropy (neglecting thermal conduction),
\beq
   \f{Ds}{Dt} = -\f{\mathcal{L}}{T},
   \label{eq:DsDt}
\seq
where  $D/Dt = \pdtext{}{t} + \gv{v}\cdot\nabla$ is the advective derivative.  
It is convenient to make this equation dimensionless.  If we introduce the primed variables $s'\equiv s/\cv$, $t'\equiv t/t_{\rm cool}$, $\mathcal{L}' \equiv \mathcal{L}/\Lambda_0$, and $T'\equiv T/T_0$, this equation becomes $Ds'/Dt' = -\mathcal{L}'/T'$ for $t_{\rm cool} = \cv T_0/\Lambda_0$ as defined in \eqref{eq:tcool}.\footnote{Whereas in linear theory, the subscript `0' denotes a quantity evaluated in the uniform background flow, here it denotes just a fiducial value, as \eqref{eq:DsDt} is a dynamical equation for how the background flow varies.}
To proceed without an excessive use of `primes', all equations from here are understood to be written in terms of these primed variables.

As mentioned by \cite{Balbus86}, applying a Lagrangian perturbation operator to \eqref{eq:DsDt} is the key 
to understanding TI anytime the flow is in a TNE state, having departed from the equilibrium curve.  
This hints at it being useful to apply a second advective derivative instead, to give
\beq
   \f{D^2s}{Dt^2} = -\f{D}{Dt}\f{\mathcal{L}}{T}.
   \label{eq:D2sDt2}
\seq
In Appendix~B, we show that for net cooling functions of the form $\mathcal{L} = \mathcal{L}(\rho,T)$, \eqref{eq:D2sDt2} is equivalent to 
\beq
   \f{D^2s}{Dt^2} = 
    N_p' \f{D\ln\rho}{Dt} - N_\rho' \f{D\ln p}{Dt},
   \label{eq:main_eqn}
\seq
where $N_p'$ and $N_\rho'$ are just the quantities from \eqref{eq:Nrho_prime} and \eqref{eq:Np_prime} expressed in dimensionless variables:
\beq
\begin{split}
N_p' &= T\left(\pd{\mathcal{L}/T}{T}\right)_p ; \\
N_\rho' &= T\left(\pd{\mathcal{L}/T}{T}\right)_\rho .
\end{split}
\label{eq:NpNrho_prime}
\seq
\eqref{eq:main_eqn} is therefore a dynamical relation that connects gas density and pressure gradients with the quantities determining the stability criteria and growth rates of condensation modes. 
Whereas in linear theory, $N_p$ and $N_\rho$ are considered constant parameters describing the background flow, here $N_p'$ and $N_\rho'$ are time-dependent and this equation can be applied to the modes themselves to understand their nonlinear evolution.  It becomes immediately clear that once the Balbus contour is crossed (i.e., once $N_p'$ changes sign), $D\ln \rho/Dt$ and $D\ln p/Dt$ must also change, implying that the linear growth regime given by \eqref{eq:linear_profiles} has ended.

Less elegant versions of \eqref{eq:main_eqn} are needed for understanding the saturation dynamics of TI in detail.  In going from \eqref{eq:D2sDt2} to \eqref{eq:main_eqn}, we have already assumed an ideal gas, making \eqref{eq:main_eqn} not yet fully simplified.  The specific entropy is $s = \ln(p/\rho^\gamma)$ up to an additive constant.  Hence, either the left hand side of \eqref{eq:main_eqn} can be written in terms of $D^2p/Dt^2$ and $D^2\rho/Dt^2$, or either $D\ln p/Dt$ or $D\ln \rho/Dt$ can be eliminated in favor of $Ds/Dt$ on the right hand side.  
Retaining both forms of the second option is the key to understanding the full saturation mechanism.
By analogy with \eqref{eq:Rdef}, we define the ratio
\beq
R' \equiv \f{N_p'}{\gamma N_\rho'},
\label{eq:Rprime}
\seq
leading to, after a bit of algebra,
\beq
\begin{split}
\f{D^2s}{Dt^2} + \f{N_p'}{\gamma} \f{Ds}{Dt} &= -N_\rho'(1-R') \f{D\ln p}{Dt}; \\
\f{D^2s}{Dt^2} + N_\rho' \f{Ds}{Dt} &= -\gamma N_\rho'(1-R') \f{D\ln \rho}{Dt}.
\label{eq:sat_eqn_s}
\end{split}
\seq 
Eliminating $Ds/Dt$ and $D^2s/Dt^2$ using \eqref{eq:DsDt} and \eqref{eq:D2sDt2}, respectively, and solving for $D\ln p/Dt$ and $D\ln \rho/Dt$, we arrive at 
\beq
\begin{split}
\f{D\ln p}{Dt} = \f{1}{N_\rho'(1-R')}\left[\f{D}{Dt}\f{\mathcal{L}}{T} + \f{N_p'}{\gamma} \f{\mathcal{L}}{T} \right] ; \\
\f{D\ln \rho}{Dt} = \f{1}{\gamma N_\rho'(1-R')}\left[\f{D}{Dt}\f{\mathcal{L}}{T} + N_\rho' \f{\mathcal{L}}{T} \right] .
\label{eq:sat_eqn}
\end{split}
\seq
The first of these final expressions reveals that sign changes of $D\ln p/Dt$ are determined by the location on the phase diagram (the value of $N_p'$, $N_\rho'$, $R'$, and $\mathcal{L}/T$) as well as by $D(\mathcal{L}/T)/Dt$, the rate of change of the instantaneous entropy production rate as measured by tracking a fluid element.  The latter quantity accounts for the kinematic motion of the fluid element that is determined by the continuity and force equations.

The second expression is interpreted similarly, but it is distinct in that $N_p'$ no longer appears in the expression in brackets.  As we show in \S{3.3}, for dynamics taking place once gas lands on the equilibrium curve, sign changes in $D\ln \rho/Dt$ are accompanied by sign changes in $D\ln p/Dt$, and this is associated with `pulsations' --- oscillations in the size of the cloud. 

\subsection{Crossing the Balbus contour}
We now apply \eqref{eq:sat_eqn} to a single condensation mode in the nonlinear phase of TI to infer the sequence of dynamical events that must take place during saturation.  We focus our analysis on the blue horizontal tracks in figure~1, corresponding to nearly isobaric evolution within a TI zone (where $N_\rho' > 0$, $N_p' < 0$, and $R' < 0$).  It is clear from figure~1 that $N_p'=0$ will occur well before the cooling gas reaches the equilibrium curve (at which point $n \approx 9\times10^{11}~{\rm cm^{-3}}$).  Evaluating the $D\ln p/Dt$ equation at $N_p' = 0$ gives
\beq 
\f{D\ln p}{Dt} = \f{1}{N_\rho'(1-R')} \f{D}{Dt}\f{\mathcal{L}}{T} .
\label{eq:Np_is_0}
\seq
This equation states that, provided $R'<1$, there can be only two possibilities once the density has increased enough to place the cool gas at the Balbus contour: 
(i) the sign of $D\ln p/Dt$ is the same as that of $D(\mathcal{L}/T)/Dt$; or
(ii) both $D\ln p/Dt$ and $D(\mathcal{L}/T)/Dt$ are zero.  We will henceforth refer to the event of $D\ln p/Dt$ passing through zero as a `pressure reversal'.  The proviso that $R' < 1$ is confirmed in figure~2.

Possibility (ii) would imply that a pressure reversal accompanies a Balbus crossing and also coincides with the time that $\mathcal{L}/T$ reaches its maximum value.
Our claim is that possibility (i) would imply that all three events are not simultaneous but that a pressure reversal is nevertheless associated with the event of gas reaching $N_p' = 0$.   
If the evolution were to exactly follow an isobar, then using the chain rule, 
\beq
\f{D}{Dt}\f{\mathcal{L}}{T} = \left( \pd{\mathcal{L}/T}{T} \right)_p \f{DT}{Dt} .
\label{eq:isobar}
\seq 
The Balbus contour is by definition where the first term in parenthesis vanishes, so in this instance there is only possibility (ii) above.  Because perfect isobaric evolution is in violation of \eqref{eq:linear_profiles}, we conclude that these events cannot be simultaneous.  The causal link has been established, however, and we expect the interval between a Balbus crossing and a pressure reversal to be a measure of (non)isobaricity.

To sort out the order in which the events occur, we can follow similar reasoning and express Balbus' instability criterion from \eqref{eq:BalbusCriterion} as
\beq 
\left[ \left(\pd{T}{t}\right)^{-1} \pd{}{t}\f{\mathcal{L}}{T} \right]_p  < 0.
\label{eq:BalbusCriterion_t}
\seq 
This expression reveals that unstable cooling gas has $\pdtext{(\mathcal{L}/T)}{t} > 0$ at any fixed position within the overdense region of a condensation mode. It therefore tells us that the sign of $D(\mathcal{L}/T)/Dt$ starts off positive during the linear growth phase.   
By \eqref{eq:linear_profiles}, $D\ln p/Dt < 0$ in the linear regime --- physically, unstable cooling gas loses pressure support because the temperature drops faster than the density can rise.  As the Balbus contour is approached, it is possible mathematically for $D\ln p/Dt$ to reach 0 before $D(\mathcal{L}/T)/Dt$ does; by \eqref{eq:sat_eqn}, $D\ln p/Dt = 0$ corresponds to 
\beq 
\f{D}{Dt}\f{\mathcal{L}}{T} = -\f{N_p'}{\gamma} \f{\mathcal{L}}{T}. 
\label{eq:DlnpDt_is_0}
\seq 
Above the Balbus contour $N_p' < 0$, giving $D(\mathcal{L}/T)/Dt > 0$ as required.  However, this order of events implies that a pressure reversal was not preceded by a sign change in $D(\mathcal{L}/T)/Dt$ or a change in stability conditions.  On this basis, we can discount this outcome except in instances where the cooling gas passes through the isochoric TI zone (see \S{3.4}).

The remaining outcome is the following sequence of events:
\begin{enumerate}
    \item The cooling rate $\mathcal{L}/T$ reaches its maximum value prior to crossing the Balbus contour, i.e. $D(\mathcal{L}/T)/Dt = 0$ in a region where $N_p' < 0$.  After this time, $D(\mathcal{L}/T)/Dt < 0$.
    \item The cooling gas reaches the Balbus contour.  Since $D(\mathcal{L}/T)/Dt < 0$ now, $D\ln p/Dt < 0$ according to \eqref{eq:Np_is_0}, meaning the sign of $D\ln p/Dt$ is still that of the linear growth phase.
    \item Prior to reaching the equilibrium curve where $\mathcal{L} = 0$, a pressure reversal takes place when  \eqref{eq:DlnpDt_is_0} is satisfied.  Because $N_p' > 0$ now, $D(\mathcal{L}/T)/Dt < 0$ by \eqref{eq:DlnpDt_is_0}, consistent with the first event.   
\end{enumerate}
We conclude that the true trigger for saturation is not crossing the actual Balbus contour corresponding to $N_p' = 0$ on the phase diagram, but rather the `local Balbus contour' seen by a comoving fluid element, $D(\mathcal{L}/T)/Dt = 0$.  In practice, however, it is likely that these first two events occur within one thermal time, making the distinction unimportant.  

\begin{figure}[h!]
\begin{center}
\includegraphics[width=0.78\textwidth]{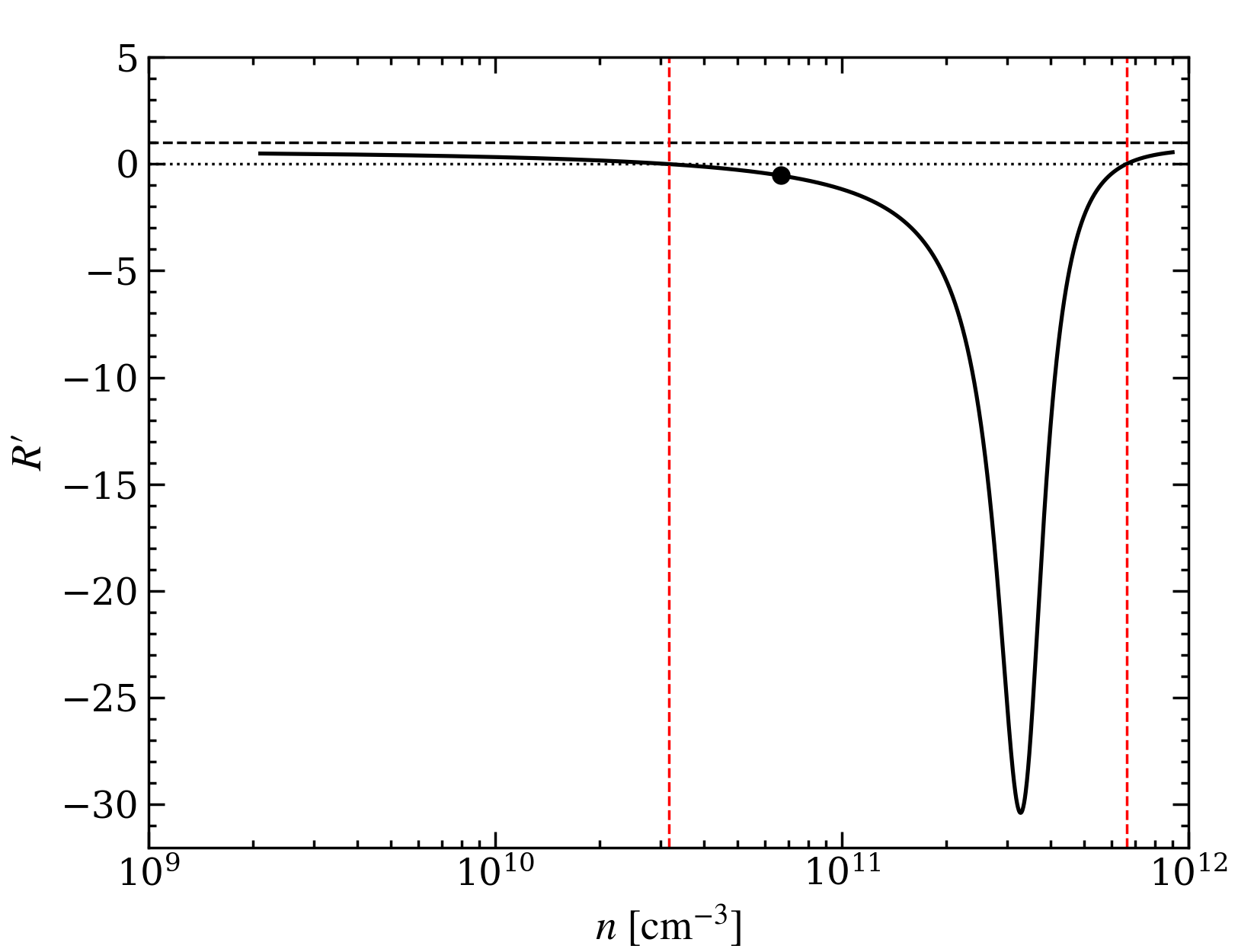}
\end{center}
\caption{Variation of the parameter $R'$ defined in \eqref{eq:Rprime} during isobaric evolution.  The red vertical lines demarcate the TI zone corresponding to the horizontal tracks in figure~1, with a black dot again marking the initial equilibrium state of the background flow.  Once a short wavelength condensation mode reaches nonlinear amplitudes, it will sample the $R'$ values given by the black curve.  This curve terminates on either end at stable ($R' > 0$) points on the equilibrium curve.  A dashed horizontal line is drawn at $R' = 1$; the property $R' < 1$ for isobaric evolution is likely a generic one for astrophysical cooling functions.
}
\label{fig:2}
\end{figure} 

\subsection{Landing on the equilibrium curve}
Figure~2 shows that $R' < 1$ during the entire path traced by the condensation mode as it isobarically approaches the equilibrium curve.  This is the expected result based on linear theory (see \S{2.1}). Sign changes in $D\ln p/Dt$ and $D\ln \rho/Dt$ are therefore controlled by the bracketed terms in \eqref{eq:sat_eqn}.  We will reserve the phrase `pressure reversal' to apply exclusively to the circumstance that $D\ln p/Dt$ changes sign due to crossing the Balbus contour.  Other options for sign changes will be referred to as `zero crossings'.  

We already mentioned that $N_p'$ does not appear in the bracketed term of the $D\ln \rho/Dt$ equation, and this means that contraction continues during and after a pressure reversal.  
Because $D(\mathcal{L}/T)/Dt < 0$ outside the TI zone after the Balbus contour is crossed, the first time that $D\ln \rho/Dt$ can reach 0 is, according to \eqref{eq:sat_eqn}, when
\beq 
\f{D}{Dt}\f{\mathcal{L}}{T} = - N_\rho'\f{\mathcal{L}}{T}. 
\label{eq:DlnrhoDt_is_0}
\seq 
This event corresponds to the density reaching its maximum value, thus marking the transition from contraction to expansion.  In other words, this marks the beginning of the contraction rebound process.  

In an isobaric case, it is likely for \eqref{eq:DlnrhoDt_is_0} to be satisfied only once, this single contraction rebound event leading directly to the cloud reaching a steady state on the equilibrium curve.   
For nonisobaric evolution, the pulsations that occur after contraction rebound, as observed in numerical simulations \citep[see WP19;][]{GronkeOh20}, correspond to cycles between subsequent zero crossings of $D\ln p/Dt$ and $D\ln \rho/Dt$, each zero crossing satisfying \eqref{eq:DlnpDt_is_0} and \eqref{eq:DlnrhoDt_is_0}.  Notice that this will involve oscillations about the equilibrium curve $\mathcal{L} = 0$.  A dedicated study of the $D\ln \rho/Dt$ equation in \eqref{eq:sat_eqn} is necessary to understand what parameters govern the pulsation damping rate. 

\subsection{Passing through the isochoric TI zone}
Notice from figure~1 that for the net cooling function given in Appendix~A, it is possible to enter a regime of isochoric TI \textit{during isobaric evolution} at just slightly higher pressures than that of the black dot.  As mentioned in \S{3.2}, this allows for a pressure reversal to occur inside the isochoric TI zone at a location where \eqref{eq:DlnpDt_is_0} can be satisfied.  The only new constraint is found by setting $N_\rho' = 0$ in \eqref{eq:sat_eqn} to give
\beq 
\f{D\ln\rho}{Dt} = -\f{1}{N_p'} \f{D}{Dt}\f{\mathcal{L}}{T} .
\label{eq:isochoric_TIzone}
\seq 
This relation must hold at the boundary of the isochoric TI zone, revealing that if density is to remain monotonically increasing upon both entering and exiting this region (throughout which $N_p' < 0$), $D(\mathcal{L}/T)/Dt$ must remain positive.  Substituting \eqref{eq:isochoric_TIzone} into the $D\ln p/Dt$ equation in \eqref{eq:sat_eqn} additionally shows that a pressure reversal will occur at this boundary if $D\ln\rho/Dt = \gamma^{-1}\mathcal{L}/T$ there.
The alternative is for $\mathcal{L}/T$ to reach its maximum value at the boundary of this zone, which would correspond to having $D\ln \rho/Dt = 0$, thereby satisfying \eqref{eq:DlnrhoDt_is_0} also.  This would imply that contraction rebound precedes the pressure reversal because $D\ln p/Dt = -\mathcal{L}/T$ by \eqref{eq:sat_eqn}, i.e. the sign of $D\ln p/Dt$ is still that of the linear growth phase.  The viability of either of these sequences of events needs to be verified numerically.

\subsection{Other applications of the new identities}
Being a general identity of nonadiabatic gas dynamics, \eqref{eq:main_eqn} may prove useful in a variety of settings.\footnote{Note this equation is more general than \eqref{eq:DsDt} because that equation neglects all effects of thermal conduction due to a heat flux vector $\gv{q}$, whereas \eqref{eq:D2sDt2}, \eqref{eq:main_eqn}, and \eqref{eq:sat_eqn_s} will still hold in regions where $\div\gv{q} = constant$.}
In particular, \eqref{eq:main_eqn}-\eqref{eq:isochoric_TIzone} are unchanged in ideal MHD, hence so should be the overall saturation mechanism.  However, the thermal evolution can differ in detail because in MHD, paths traced on the phase diagram depend on the field strength \citep[][WP19]{Bottorff2000}.  The linear theory of TI in MHD has recently been revisited by \citet{Claes19}, and it would be interesting to determine the plasma beta at which unstable condensation modes no longer carry out the sequence of events listed in \S{3.2}.

Another obvious application is to the connection between TI and convective instability in stratified atmospheres \citep[see][]{Balbus89,Binney09,BalbusPotter}.  To show that \eqref{eq:main_eqn} is a generalization of results derived in that context, it is sufficient to examine one of the identities in \eqref{eq:sat_eqn_s} in a steady state. Taking $D/Dt \rightarrow \gv{v}\cdot\nabla$ gives, for the $D\ln p/Dt$ equation,
\beq 
\gv{v}\cdot\nabla(\gv{v}\cdot\nabla s) + \f{N_p'}{\gamma} \gv{v}\cdot\nabla s = -N_\rho'(1-R') \gv{v}\cdot\nabla \ln p .
\seq
Noting that $N_\rho'(1-R') = N_\rho' - N_p'/\gamma$ by \eqref{eq:Rprime}, we now also assume the background state of the gas is in thermal equilibrium with $\mathcal{L}=0$.  This reduces the quantity $N_\rho' - N_p'/\gamma$ to $T[(\pdtext{\mathcal{L}}{T})_\rho - \gamma^{-1}(\pdtext{\mathcal{L}}{T})_p]$, which is simply related to the quantity $(\pdtext{\mathcal{L}}{T})_s$ by the thermodynamic identity \citep[see e.g.][]{Balbus95},
\beq 
\f{\gamma-1}{\gamma} \left(\pd{\mathcal{L}}{T}\right)_s = \left(\pd{\mathcal{L}}{T}\right)_\rho - \f{1}{\gamma} \left(\pd{\mathcal{L}}{T}\right)_p,
\seq
bringing us to 
\beq 
\gv{v}\cdot \left[\nabla(\gv{v}\cdot\nabla s) + \f{1}{\gamma} \left(\pd{\mathcal{L}}{T}\right)_p \nabla s + \f{\gamma-1}{\gamma} \left(\pd{\mathcal{L}}{T}\right)_s \nabla \ln p \right] = 0 .
\seq
If we now apply an Eulerian perturbation operator to this equation, the result is a product rule giving, schematically, $\delta \gv{v} \cdot [\ldots] + \gv{v}\cdot \delta[\ldots] = 0$.  As a last step, we assume the background atmosphere is static and spherically symmetric, leaving just the bracket attached to $\delta\gv{v}$, which simplifies to 
\beq 
\left(\pd{\mathcal{L}}{T}\right)_p \pd{s}{r} = -(\gamma-1)\left(\pd{\mathcal{L}}{T}\right)_s \pd{\ln p}{r}.
\seq
This equation is identical to Eq.~(96) in \citet{BalbusPotter}, who obtained it through a more involved Lagrangian perturbation analysis.  As they explain, this relation implies that TI will be accompanied by convective instability in an unstable static atmosphere.

In complicated flows, TI zone boundary crossings are likely common if not inevitable events during thermal evolution.
Pressure reversals could therefore also play a role in mediating fragmentation caused by flow interactions.  To provide an illustrative example, first note that a pre-existing cloud put in `by hand' as initial conditions in nonadiabatic cloud-wind simulations, as well as the fragments that appear as the cloud's surface layers are disrupted, all occupy the same parameter space on a phase diagram as a cloud that formed from TI.  Referring to the tracks in figure~1, as this gas changes its temperature upon interacting with the wind, the cold phase gas may repeatedly enter and exit the TI zone.  A less disruptive example where crossing the Balbus contour may initiate the underlying fragmentation process is the `tearing' mechanism identified by \citet{JenningsLi}.
By following individual fluid elements and applying \eqref{eq:sat_eqn}, it should be possible to decipher the causal relationship between forces and the overall thermal evolution.

Yet another topic concerns the dynamical role of pulsations in a regime of isochoric TI.  Pulsations correspond to a change in sign of $D\ln\rho/Dt$, as discussed in \S{3.3}.  We have not yet studied nonisobaric evolution in a regime of isochoric TI using numerical simulations, and it is unclear if other groups have either.  It is not standard practice to calculate the value of $R_\lambda$ in \eqref{eq:Rdef} or $R'$ in \eqref{eq:Rprime} that characterizes any particular TI regime.  This is especially relevant for recent work on cloud coalescence instability: \cite{GronkeOh22} claim that this instability is aided by overstable acoustic modes, but such modes only exist when the isochoric or isentropic instability criterion is satisfied.  Because Gronke \& Oh do not indicate which instability regime they consider, their claim of unstable acoustic modes being the agent that increases the coalescence rate is questionable.  In our work on this instability \citep{Waters19b}, we mentioned that 
coalescence can potentially be very fast: its rate can reach the rate set by the dynamical timescale provided the flow is continually subject to thermal disturbances, 
for this (re)excites pulsations and their accompanying advective flows that mediate the entire merger process.

\section{Summary and discussion of controversial claims}
In previous studies we have shown that plotting the Balbus contour (the boundary of the TI zone on a phase diagram) 
is an essential diagnostic for understanding dynamical TI \citep{Barai12,Dannen20,Waters21,Waters22}, 
the counterpart to local TI when the background flow gradients are nonzero --- and the only type of TI encountered in global accretion flow or outflow simulations.  Here we have established that the Balbus contour plays an important dynamical role for local TI also.  
Specifically, \eqref{eq:sat_eqn} reveals the mechanism by which TI saturates:
whenever this contour is crossed, a sequence of events unfolds that causes 
$D p/D t$ to undergo a change in sign.  This marks the end of the exponential growth phase, 
i.e., crossing the Balbus contour causes the density, velocity, and pressure profiles to no longer 
resemble the solution to the linearized equations given in \eqref{eq:linear_profiles}. 

The saturation process is initiated in gas undergoing cooling.  The pressure reversal implied 
by a change in sign of $D p/D t$ indicates that the condensation can start gaining pressure support; 
at a fixed location within the condensation, $-\nabla p$ changes from pointing inward to outward 
and thus begins to halt further contraction. The confining gas, meanwhile, continues to occupy 
the TI zone and undergo runaway heating because the hotter gas is less dense and therefore heats up 
at a lower rate than the condensation cools down.  This, in turn, means that the confinement pressure 
continues to rise during and after contraction rebound.  Once the hot phase gas rises above 
the Balbus contour and undergoes a pressure reversal, its pressure switching from increasing to decreasing 
with time, the reduction in the confinement pressure aids further expansion of the condensation until, 
in an isobaric case, a steady state is reached. 

In a nonisobaric case, i.e., for wavelengths large enough that the condensation velocity becomes 
a significant fraction of the ambient sound speed, pulsations will set in after contraction rebound.  
Pressure oscillations first arise through the nonlinear interaction between the condensation and the sound waves excited when each fluid element undergoes a pressure reversal (which occurs at different times for different locations in the cloud core).
To emphasize this last point, we reiterate that in an isobaric case,
mechanical equilibrium is closely maintained, whereas in a nonisobaric case it is temporarily lost.
There is a build-up of strong gas pressure forces, so that by the time TI staturates and 
most of the gas is thermally stable (with the cold gas nearly in thermal equilibrium again), the mechanical state is still far from equilibrium.  Therefore, the next phase of the evolution is dominated by pressure waves 
that eventually bring the system to mechanical equilibrium.
The significance of the gas crossing the Balbus contour is in it being associated with 
the very first qualitative change in the gas behavior, a precursor to a chain of events resulting 
in a very dynamic evolution of the system to a new thermal and mechanical equilibrium.

\subsection{Shattering versus splattering}
In \S{2}, we did not draw a distinction between `nonisobaric evolution' as it relates to TI and nonisobaric dynamics more generally.  The latter is simply the tendency for gas with $t_{\rm cool} \lesssim t_{\rm cross}$ to undergo oscillations in response to a thermal disturbance, and the cloud coalescence simulations we presented in \citet{Waters19b} provide an example of this.  The former refers to the nonisobaric dynamics accompanying the saturation of TI: as the wavelength of a condensation mode increases, nonisobaric evolution involves increasingly strong pulsations following contraction rebound.  Extreme nonisobaric behavior is characterized by `splattering', a term we introduced to indicate a contraction rebound so strong that the cloud undergoes self-fragmentation.  

The `shattering' hypothesis by contrast, described by \citet{McCourt18} as there being a tendency for highly nonisobaric clouds to undergo spontaneous self-fragmentation, as opposed to breakup caused by a dynamical response to contraction, has not held up to scrutiny \citep{GronkeOh20,JenningsLi,Das21,Farber22}.  As we explained in \citetalias{Waters19a} and again here in more detail (see \S{2.3}), \citet{McCourt18}'s simulations, with initial conditions having a spectrum of perturbations, can be understood as a clear example of `isobaric takeover': short wavelength perturbations form many isobaric condensations within a much larger pre-existing cloud.  In other words, those simulations do not follow self-fragmentation of one entity but rather the evolution of a nonisobaric cloud serving as the background flow within which the shortest wavelength condensation modes (that have the fastest growth rates) can form clouds and interact.  \citet{BurkertLin} described such an outcome, claiming that the end result is indistinguishable from a fragmentation process.  While we disagree with this, especially on the grounds that isobaric takeover will lead to coalescence (the opposite of fragmentation), even accepting the claim does not invalidate our takeaway point, which is that `splattering' refers to a definite mechanism leading to self-fragmentation (namely, contraction rebound) while `shattering' does not.  

Despite our pointing out in WP19 that \citet{McCourt18}'s simulations are a demonstration of isobaric takeover, the term `shattering' is still regularly invoked when describing the appearance of small-scale cloud fragments in wind-cloud or shock-cloud interaction simulations that include radiative cooling \citep[e.g.,][]{Sparre20,Banda21,Bustard22,Jennings23}.  The use of the term here is likely because the size of these fragments appears to be the cooling length evaluated in the cold phase gas, denoted $\min(\lambda_{\rm cool})$, which is the isobaric length scale identified by \citet{McCourt18}.  A cloud fragment with characteristic size $d_c$ has an associated crossing time $t_{\rm cross} =d_c/c_s$ and can be said to be isobaric if $t_{\rm cross} \lesssim t_{\rm cool}$. Equivalently, because $\lambda_{\rm cool} = c_s t_{\rm cool}$ is the length scale over which sound waves can effectively communicate thermal disturbances, isobaric cloud sizes satisfy $d_c \lesssim \lambda_{\rm cool}$, meaning that opposite sides of the cloud remain in sonic contact.  The finding that interactions with a wind cause fragments to reach this length scale is not an unexpected outcome, hence this occurrence should not be confused with the regions of the cloud \textit{not interacting} with the wind undergoing fragmentation, which is what `shattering' would entail.  
Moreover, it is crucial to note that in \citet{Jennings23}'s simulations, these isobaric fragments all evaporated in the runs with thermal conduction, consistent with our result that $\min(\lambda_{\rm cool})$ is generically smaller than the Field length ($\lambda_F$) evaluated in the warm phase gas \citepalias[see][]{Waters19a}.  Thus, while $\min(\lambda_{\rm cool})$ is a characteristic size scale for cloud fragments, it is not a physically relevant one unless thermal conduction is highly suppressed.

While `splattering' refers exclusively to a highly nonisobaric regime of TI, the pulsation behavior it relates to could permit the characteristic size of cloud fragments to significantly exceed the scale $\min(\lambda_{\rm cool})$ in nonadiabatic wind-cloud interaction simulations.  These fragments are expected to become larger upon increasing the wind temperature (while keeping the wind pressure the same), based on the following reasoning.  The descriptors `isobaric' and `nonisobaric' have thus far been used for both the size and the behavior of the cold phase gas, but they also apply to the confining warm phase gas.  A parcel of warm phase gas is isobaric on scales $d_w < \max(\lambda_{\rm cool})$, where $\max(\lambda_{\rm cool})$ denotes evaluating the cooling length at the warm phase temperature.  This gas can easily remain isobaric, depending on how much hotter it is compared to the cold phase, and can therefore be effective in dampening the pulsations of nonisobaric clouds when $\max(\lambda_{\rm cool}) \gg d_c$.  In other words, two equal size clouds with $d_c \gg \min(\lambda_{\rm cool})$ (and with the same pressure) embedded in different temperature environments, will not exhibit the same nonisobaric behavior when subjected to the same thermal disturbance.\footnote{An idealized version of this thought experiment is `the pulsating sphere', a well known problem in acoustics \citep[e.g.][]{Devaud13}.} The pulsation damping rate is expected to be larger in the higher temperature environment, making that cloud effectively less nonisobaric, and therefore less prone to fragmentation when interacting with a shearing flow.  Even if this effect is significant, the fragments would still evaporate when including isotropic thermal conduction because $\lambda_F \propto T^{11/4}/p$ (under Spitzer conductivity and fixed gas pressure $p$), while $\max(\lambda_{\rm cool}) \propto T^{5/2}/p$.  Higher wind temperatures might allow larger fragments to survive when thermal conduction is anisotropic, however.  

\subsection{Thermal instability versus thermal nonequilibrium}
In the solar physics literature, the relation between TNE and TI is explained by reference to a TNE-TI cycle that describes observed phenomenology \citep[see e.g.,][]{Antolin22}, specifically the dynamics taking place in coronal loops (magnetic flux tubes that extend from the chromosphere into the corona but are rooted to the solar surface at both ends).  TNE itself has been described as a process that occurs when the heating rate profile in a loop drops off sharply with height and is unchanging with time, while the cooling rate remains a local quantity, making it possible for there to be no `nearby' thermal equilibrium state (in $\rho$-$T$ phase space) where cooling can reach a balance with heating \citep{Klimchuk19}.  Both evaporative and bulk flows can occur in response to pressure changes during a TNE cycle.  

\citet{Klimchuk19} claimed that it is meaningless to consider TI occuring under TNE conditions on the (false) premise that diagnosing TI requires the background flow to be in a steady state.
We stress that \citet{Balbus86}'s stability criterion superseded Field’s criterion for the very purpose of applying to TNE conditions.  For there to be a consensus on this issue, the concept of TNE as given in the preceding paragraph would need to be generalized to adopt the definition we used in \S{2.4}, namely that TNE is simply the circumstance that a flow departs from the equilibrium curve, meaning $\mathcal{L} \neq 0$.  
We should furthermore specify that the bulk flow is subjected to $\mathcal{L} \neq 0$ on dynamical timescales to distinguish TNE from \textit{thermal misbalance}, which simply refers to there being fluctuations about $\mathcal{L} = 0$ on very short timescales due to acoustic compression and expansion cycles that result in the damping of MHD waves \citep[e.g.][]{Kolotkov19,Kolotkov21}. 

Under TNE conditions then, the criterion for entropy modes to be thermally unstable is given by \eqref{eq:BalbusCriterion}, which expands to
\beq 
\left(\pd{\mathcal{L}}{T}\right)_p  < \f{\mathcal{L}}{T}.
\label{eq:BalbusCriterion2}
\seq 
This reduces to the isobaric instability criterion of \citet{Field65} when the gas reaches thermal equilibrium with $\mathcal{L} = 0$.  In the astrophysics literature, the canonical example of a plasma that violates either criterion is one with \citep[e.g.,][]{McCourt12,MP13,BalbusPotter}
\beq 
\mathcal{L} = A \rho T^d - B,
\seq 
where $A$, $B$, and $d$ are constants satisfying $A > 0$, $B \geq 0$, and $d < 1$. 
This is, however, the same example used by \citet{Klimchuk19} in an attempt to illustrate a condensation forming under TNE conditions in the absence of TI.\footnote{We note that \citet{Klimchuk19} introduced a (half-wavelength) entropy mode into his initial conditions by the choice of a density profile given by \eqref{eq:linear_profiles} with $A = 0.01$.}

Because gas can be in TNE due to a variety of causes, a further distinction needs to be drawn between local TI and dynamical TI.  The theory of local TI, as summarized in \S{2.1}-\S{2.2}, is not confined to a homogeneous gas; it applies whenever the so-called \textit{local approximation} holds, i.e., on length scales $\Delta x$ over which $\Delta x \ll \lambda_q$, where $\lambda_q \equiv |\nabla \ln q|^{-1}$ is the scale length for the gradient of the background flow quantity $q$ to be significant.  Dynamical TI is the situation when the growth rates predicted by local TI cease to be valid because $\Delta x \sim \lambda_q$ for at least one of the relevant flow variables among $q = (\rho, \gv{v}, p, T)$. 

Given the above considerations, the background flow for local TI can be an initially uniform region that is in an evolving equilibrium state or in a TNE state.  As discussed in \S{2.4}, the various isochoric instability regimes are associated with gas starting off in TNE, as the entire isochoric TI zone is far from the equilibrium curve.  In \citet{Proga22}, we provided a qualitative example of a column of gas starting off thermally stable but becoming unstable to local TI upon following a time-dependent equilibrium curve.  In dynamical TI, by contrast, a flow can be attempting to evolve along the equilibrium curve to maintain $\mathcal{L} = 0$, but the timescale associated with adiabatic cooling can become shorter than $t_{\rm cool}$, thereby causing the flow to tend toward a steady state at some position off the equilibrium curve where $\mathcal{L} \neq 0$.  

If the coronal rain observed in coronal loops is attributable to the saturation of condensation modes, then due to the presence of flows and a chromosphere-corona transition region in these loops, the condensations are a clear instance of dynamical TI.  Rather than the black dot in figure~1, which represents a constant density initial condition appropriate for local TI, let us consider instead a possible initial coronal loop TNE state having a density spanning the range $n\approx 10^{11}-10^{12}~{\rm cm^{-3}}$ and a nearly flat pressure profile that places it below the equilibrium curve in a region of net heating (with $\mathcal{L} < 0$).  The plasma will approach the equilibrium curve with time, but the flow traversing a converging (diverging) portion of the loop will undergo slight adiabatic heating (cooling), making the phase diagram `tracks' of this flow spread vertically at different rates even if the heating rate is uniform.  The plasma with $n \gtrsim 1.5\times10^{11}~{\rm cm^{-3}}$ is very close to the stable cold branch of the equilibrium curve and will therefore not be subject to runaway heating.  However, the stable equilibrium state of the plasma with $n < 1.5\times10^{11}~{\rm cm^{-3}}$ is the hot branch where $n \lesssim 10^{10}~{\rm cm^{-3}}$.  If pressure equilibrium can be closely maintained, the bulk of the coronal region plasma may approach this branch without ever entering the TI zone. 

This is, in essence, a TNE runaway heating scenario that leads to a multi-temperature plasma without ever invoking TI.  We are therefore agreeing with the overall premise of \citet{Klimchuk19}, but our example debunks his assertion that ``the physics that governs the thermal runaway in a TNE loop is
equivalent to the physics that governs the thermal runaway in a [thermally] unstable equilibrium loop''.
To examine condensation formation occurring under TNE conditions because of TI, we could alternatively imagine initial conditions that place the gas above the equilibrium curve in figure~1 in a region of net cooling.  Here we would have to confront both bulk runaway cooling and the saturation of unstable entropy modes, but we have hopefully made it clear that these are distinct processes. 

Either coronal loop scenario is complicated by the fact that equilibrium curves are in general time-dependent, i.e., the contour $\mathcal{L} = 0$ can change with time according to how the radiation field or other background sources of heating evolve.  If this evolution occurs on timescales shorter than $t_{\rm cool}$, the plasma cannot maintain $\mathcal{L} = 0$ even if it was able to settle on the equilibrium curve, and it will thus re-enter TNE.  The heating and cooling rates used for illustrative purposes in this paper allow studying this situation.  Namely, the equilibrium curve in figure~1 models a column of gas being irradiated by both a thermal and non-thermal source of X-rays, and the two parameters $T_{C,h}$ and $f_{h}$ (controlling the temperature and flux of the non-thermal photons, respectively) can be made time-dependent to vary the effective Compton temperature, which controls the shape of the equilibrium curve (see Appendix~A).

We have recently worked on several other applications of dynamical TI that reveal the interplay between TI and TNE.  In \citet{Waters21}, we showed that the multiphase radial outflow solutions discovered in 1D by \citet{Dannen20} can reach a steady state, permitting a formal stability analysis of an outflow occupying parameter space with $\mathcal{L} \neq 0$.  There we also investigated the interconnection between TI and runaway heating in 2D simulations of accretion disk winds driven by external irradiation from X-rays, showing that it is possible to identify the presence of local TI even in a time-dependent, turbulent flow by plotting the `tracks' of individual streamlines on the phase diagram to see if they pass through a TI zone.  Finally, in \citet{Waters22}, we addressed a different point raised by \citet{Klimchuk19}.  Referring to the stratified flow within a magnetic flux tube, he posed the question, ``if a perturbation grows, does it have time to reach a substantial amplitude before it is carried to the chromosphere by the flow?''
We showed how to calculate whether or not outflowing entropy modes have time to saturate when they sample a time-dependent growth rate as they `fly through' a TI zone. 

\appendix
\section{Appendix A: net cooling function}
Arriving at the expressions used to make figure~1 requires defining a few variables specific to RHD.  
In a gray treatment of the radiation, the mean intensity at a given location is $J = (4\pi)^{-1} \int I \, d\Omega$, where $I$ is the frequency-integrated specific intensity.  We assumed a point source of radiation consisting of two underlying spectral components: $\Js$, a soft source of ionizing photons that corresponds to a blackbody above $10^4~\rm{K}$, and $\Jh$, a hard source with a high energy non-thermal spectrum.  The ionizing flux that irradiates the face of a distant slab is $F_{\rm ion} = 4\pi(\Js + \Jh)$. The ionization parameter, defined as $\xi \equiv 4\pi F_{\rm ion}/n$, becomes $\xi = (4\pi)^2 (\Js + \Jh)/n$.  
The energy density of the radiation is related to the mean intensity by $E_r = (4\pi/c) J$, so we also have $\Ers = (4\pi/c) \Js$ and $\Erh = (4\pi/c) \Jh$.
If we then specify $\Jh$ in terms of $\Js$ through the parameter $f_{\rm h} \equiv \Jh/\Js$, we can parameterize the radiation energy density as
\beq 
\begin{split}
    \Ers &= \f{\xi_0 n_0}{4\pi c}\f{1}{1+f_{\rm h}}, \\
    \Erh &= f_{\rm h} \Ers,
\end{split}
\label{eq:Erdef}
\seq 
where $\xi_0$ and $n_0$ are the free parameters setting the boundary conditions at the face of the slab.
 
It is assumed that the thermal radiation interacts with matter through both free-free absorption and Compton scattering.  The former is modeled using a Kramer's opacity; specifically, the Planck mean absorption opacity is $\kappa_{aP} = A_{\rm ff}\rho T^{-3.5}$ and we set $A_{\rm ff} = 5.19\times 10^{24}{\rm cm^5 g^{-2} K^{3.5}}$.  The non-thermal radiation only interacts via Compton scattering.  In the optically thin limit, attenuation caused by these interactions is negligible, so that $E_r = \Ers + \Erh$ is a constant.  The optically thin cooling function that results is
\beq
\mathcal{L} = c A_{\rm ff} \rho\, T^{-3.5}(aT^4 - \Ers) + c A_C E_r (T-T_{\rm C,eff}),
\label{eq:mathcalL}
\seq 
where $A_C = 4\,k_B\kappa_e/(m_e c^2) \approx 2.683\times 10^{-10}{\rm cm^2 g^{-1} K^{-1}}$ and $T_{\rm C,eff}$ is the effective Compton temperature given by
\beq
T_{\rm C,eff} = \f{T_{r,{\rm s}} + T_{\rm C,h}}{1 + f_{\rm h}}.
\seq 
Here, $T_{r,{\rm s}} \equiv (\Ers/a)^{1/4}$ is the thermal radiation temperature, while 
$T_{\rm C,h}$ is the non-thermal radiation temperature (the Compton temperature corresponding to $\Jh$; see \citet{Proga22}).  

\subsection{Analytic expressions for the equilibrium curve and TI zone boundaries}
The equilibrium curve in figure~1 was constructed by setting $\mathcal{L}=0$ in \eqref{eq:mathcalL} and solving for $n = \rho/\bar{m}$ (assuming $\bar{m} = 0.6 m_p$).  We find
\beq
n_{\rm eq}(T) = \f{A_C}{\bar{m}\,A_{\rm ff}} \f{E_r(T_{\rm C,eff}-T) }{a T^4 - \Ers} \,T^{3.5}.
\seq
Given an array of temperature values called $T_{\rm vals}$, we form an array $n_{\rm vals} = n_{\rm eq}(T_{\rm vals})$ and then plot $n_{\rm vals} T_{\rm vals}$ vs. $n_{\rm vals}$.  We chose values $n_0 = 10^{11}~{\rm cm}^{-3}$, $\xi_0 = 2\times 10^{3}~{\rm erg\,cm\,s^{-1}}$, $T_{\rm C,h} = 10^8~{\rm K}$, and $f_{\rm h} = 0.1$.

To evaluate $N_p'$, we need an expression for $\mathcal{L}/T$ with $\rho$ replaced by $p$ in \eqref{eq:mathcalL}:
\beq
\f{\mathcal{L}}{T} = c\f{\bar{m} A_{\rm ff}}{k_B} p\, T^{-5.5}(aT^4 - \Ers) + c A_C E_r \left(1-\f{T_{\rm C,eff}}{T}\right).
\label{eq:mathcalL_p}
\seq 
The isobaric temperature derivative is then
\beq\left[\pd{(\mathcal{L}/T)}{T}\right]_p  =  -\f{3}{2} c\,A_{\rm ff} \rho\, T^{-5.5}\left(aT^4 - \f{11}{3} \Ers\right)
 + c A_C E_r \f{T_{\rm C,eff}}{T^2}.
\seq
Setting this derivative to $0$ and solving for $n = \rho/\bar{m}$ gives an analytic expression for the Balbus contour:
\beq
n_{\rm B}(T) = \f{2}{3}\f{A_C}{\bar{m}\,A_{\rm ff}} \f{E_r T_{\rm C,eff} }{a T^4 - (11/3)\Ers} \,T^{3.5}.
\seq
Similarly, the isochoric temperature derivative is 
\beq\left[\pd{(\mathcal{L}/T)}{T}\right]_\rho  =  -\f{1}{2} c\,A_{\rm ff} \rho\, T^{-5.5}\left(aT^4 - 9 \Ers\right)
 + c A_C E_r \f{T_{\rm C,eff}}{T^2},
\seq
and the corresponding zero contour is 
\beq
n_{\rm isochoric}(T) = \f{2 A_C}{\bar{m}\,A_{\rm ff}} \f{E_r T_{\rm C,eff} }{a T^4 - 9\Ers} \,T^{3.5}.
\seq
As with $n_{\rm eq}$, arrays of both $n_{\rm B}$ and $n_{\rm isochoric}$ are formed by evaluating each expression using $T_{\rm vals}$.  

\section{Appendix B: derivation of Equation (18)} 
We begin by examining $Df/Dt$ for functions of the form $f = f(\rho,T)$.
Only the thermodynamic derivatives $(\pdtext{f}{\rho})_T$ and $(\pdtext{f}{T})_\rho$ enter into partial derivatives of $f$:
\beq
   \pd{f}{t} =
   \left(\pd{f}{\rho}\right)_T \pd{\rho}{t} + 
   \left(\pd{f}{T}\right)_\rho \pd{T}{t};
\seq
\beq
   \nabla f =
   \left(\pd{f}{\rho}\right)_T \nabla\rho + 
   \left(\pd{f}{T}\right)_\rho \nabla T.
\seq
It is therefore clear that
\beq
   \f{Df}{Dt}  =
   \left(\pd{f}{\rho}\right)_T \f{D\rho}{Dt} + 
   \left(\pd{f}{T}\right)_\rho \f{DT}{Dt}.
   \label{eq:DfDt}
\seq
By the thermodynamic identity,
\beq 
\left(\pd{f}{T}\right)_\rho - \left(\pd{f}{T}\right)_p 
= \f{\rho}{T} \left(\pd{f}{\rho}\right)_T, 
\seq
we can eliminate $(\pdtext{f}{\rho})_T$ in \eqref{eq:DfDt} in favor of $(\pdtext{f}{T})_p$ and $(\pdtext{f}{T})_\rho$, leading to
\beq
   \f{Df}{Dt}  =
   T\left(\pd{f}{T}\right)_\rho \left[ \f{D\ln\rho}{Dt} +  \f{D\ln T}{Dt} \right]
   -T \left(\pd{f}{T}\right)_p \f{D\ln\rho}{Dt}
\seq
after some rearrangement.  The term in brackets is simply $D\ln p/Dt$ when assuming an ideal gas ($D\ln T/Dt = D\ln p/Dt - D\ln\rho/Dt$).
Taking $f = \mathcal{L}/T$ brings us to our final expression,
\beq
   \f{D}{Dt}\f{\mathcal{L}}{T}  =
   T\left(\pd{\mathcal{L}/T}{T}\right)_\rho \f{D\ln p}{Dt}
   -T \left(\pd{\mathcal{L}/T}{T}\right)_p \f{D\ln\rho}{Dt}.
   \label{eq:DLoverTDt}
\seq
Upon defining $N_p' = T \left[\pdtext{(\mathcal{L}/T)}{T}\right]_p$, $N_\rho' = T \left[\pdtext{(\mathcal{L}/T)}{T}\right]_\rho$, and substituting \eqref{eq:DLoverTDt} into Eq.~(17), we immediately arrive at \eqref{eq:main_eqn}.

\section*{Conflict of Interest Statement}

The authors declare that the research was conducted in the absence of any commercial or financial relationships that could be construed as a potential conflict of interest.

\section*{Author Contributions}
TW derived the identities in \S{3} while working as a postdoc in DP's group at UNLV.  TW wrote the first draft of the manuscript after several discussions with DP regarding the recent literature.  Together, TW and DP analyzed the new identities to understand the saturation mechanism of TI and finalized the text. 

\section*{Funding}
Support for this work was provided by the National Aeronautics and Space Administration under TCAN grant 80NSSC21K0496.

\section*{Acknowledgments}
TW acknowledges funding support from Hui Li to attend the conference entitled \href{https://web.cvent.com/event/7b4be254-fb71-4308-b78d-0a8de242a017/summary?environment=P2}{\textit{AGN Santa Fe: where are the objects in AGN disks?}}, where this work was completed.

\bibliographystyle{Frontiers-Harvard} 
\bibliography{progalab-shared}
\end{document}